\documentclass[aip,reprint]{revtex4-1}
\usepackage{amsmath,amssymb}
\usepackage{graphicx}
\usepackage{dcolumn}
\usepackage{bm}
\usepackage{subfigure}
\usepackage{pifont} 
\usepackage{siunitx}

\usepackage[utf8]{inputenc}
\usepackage[T1]{fontenc}
\usepackage{mathptmx}
\usepackage{etoolbox}

\makeatletter
\def\@email#1#2{%
 \endgroup
 \patchcmd{\titleblock@produce}
  {\frontmatter@RRAPformat}
  {\frontmatter@RRAPformat{\produce@RRAP{*#1\href{mailto:#2}{#2}}}\frontmatter@RRAPformat}
  {}{}
}%
\makeatother
\begin{document}

\preprint{AIP/123-QED}

\title{Tunable coupling of a quantum phononic resonator to a transmon qubit with flip-chip architecture}
\author{Xinhui Ruan}
\thanks{These authors contributed equally to this work.}
\affiliation{
Beijing National Laboratory for Condensed Matter Physics,\\
Institute of Physics,Chinese Academy of Sciences,Beijing 100190,China
}
\affiliation{
Department of Automation, Tsinghua University, Beijing 100084, P. R. China
}
\affiliation{
	Key Laboratory of Low-Dimensional Quantum Structures and Quantum Control of Ministry of Education, Key Laboratory for Matter Microstructure and Function of Hunan Province, Department of Physics and Synergetic Innovation Center for Quantum Effects and Applications, Hunan Normal University, Changsha 410081, People’s Republic of China
}
\author{Li Li}
\thanks{These authors contributed equally to this work.}
\affiliation{
	Beijing National Laboratory for Condensed Matter Physics,\\
	Institute of Physics,Chinese Academy of Sciences,Beijing 100190,China
}
\affiliation{
School of Physical Sciences, University of Chinese Academy of Sciences, Beijing 100049, China
}
\author{Guihan Liang}
\affiliation{
	Beijing National Laboratory for Condensed Matter Physics,\\
	Institute of Physics,Chinese Academy of Sciences,Beijing 100190,China
}
\affiliation{
School of Physical Sciences, University of Chinese Academy of Sciences, Beijing 100049, China
}
\author{Silu Zhao}
\affiliation{
	Beijing National Laboratory for Condensed Matter Physics,\\
	Institute of Physics,Chinese Academy of Sciences,Beijing 100190,China
}
\affiliation{
School of Physical Sciences, University of Chinese Academy of Sciences, Beijing 100049, China
}
\author{Jia-heng Wang}
\affiliation{School of Integrated Circuits,Tsinghua University, Beijing 100084, China
}
\author{Yizhou Bu}
\affiliation{
	Beijing National Laboratory for Condensed Matter Physics,\\
	Institute of Physics,Chinese Academy of Sciences,Beijing 100190,China
}
\affiliation{
School of Physical Sciences, University of Chinese Academy of Sciences, Beijing 100049, China
}
\author{Bingjie Chen}
\affiliation{
	Beijing National Laboratory for Condensed Matter Physics,\\
	Institute of Physics,Chinese Academy of Sciences,Beijing 100190,China
}
\affiliation{
School of Physical Sciences, University of Chinese Academy of Sciences, Beijing 100049, China
}
\author{Xiaohui Song}
\affiliation{
	Beijing National Laboratory for Condensed Matter Physics,\\
	Institute of Physics,Chinese Academy of Sciences,Beijing 100190,China
}
\author{Xiang Li}
\affiliation{
	Beijing National Laboratory for Condensed Matter Physics,\\
	Institute of Physics,Chinese Academy of Sciences,Beijing 100190,China
}
\affiliation{
School of Physical Sciences, University of Chinese Academy of Sciences, Beijing 100049, China
}
\author{He Zhang}
\affiliation{
	Beijing National Laboratory for Condensed Matter Physics,\\
	Institute of Physics,Chinese Academy of Sciences,Beijing 100190,China
}
\affiliation{
School of Physical Sciences, University of Chinese Academy of Sciences, Beijing 100049, China
}
\author{Jinzhe Wang}
\affiliation{
	Beijing National Laboratory for Condensed Matter Physics,\\
	Institute of Physics,Chinese Academy of Sciences,Beijing 100190,China
}
\affiliation{
School of Physical Sciences, University of Chinese Academy of Sciences, Beijing 100049, China
}
\author{Qianchuan Zhao}
\affiliation{
Department of Automation, Tsinghua University, Beijing 100084, P. R. China
}
\author{Kai Xu}
\affiliation{
	Beijing National Laboratory for Condensed Matter Physics,\\
	Institute of Physics,Chinese Academy of Sciences,Beijing 100190,China
}
\affiliation{
	Hefei National Laboratory, Hefei 230088, China
}
\author{Heng Fan}
\affiliation{
	Beijing National Laboratory for Condensed Matter Physics,\\
	Institute of Physics,Chinese Academy of Sciences,Beijing 100190,China
}
\affiliation{
	School of Physical Sciences, University of Chinese Academy of Sciences, Beijing 100049, China
}
\affiliation{
	Hefei National Laboratory, Hefei 230088, China
}
\author{Yu-xi Liu}
\affiliation{School of Integrated Circuits,Tsinghua University, Beijing 100084, China
}
\author{Jing Zhang}
\affiliation{School of Automation Science and Engineering, Xi’an Jiaotong University, Xi’an 710049, China}
\affiliation{MOE Key Lab for Intelligent Networks and Network Security, Xi’an Jiaotong University, Xi’an 710049, China}

\author{Zhihui Peng}
\email{zhihui.peng@hunnu.edu.cn}
\affiliation{
	Key Laboratory of Low-Dimensional Quantum Structures and Quantum Control of Ministry of Education, Key Laboratory for Matter Microstructure and Function of Hunan Province, Department of Physics and Synergetic Innovation Center for Quantum Effects and Applications, Hunan Normal University, Changsha 410081, People’s Republic of China
}
\affiliation{
	Hefei National Laboratory, Hefei 230088, China
}
\author{Zhongcheng Xiang}
\email{zcxiang@iphy.ac.cn}
\affiliation{
	Beijing National Laboratory for Condensed Matter Physics,\\
	Institute of Physics,Chinese Academy of Sciences,Beijing 100190,China
}
\affiliation{
	School of Physical Sciences, University of Chinese Academy of Sciences, Beijing 100049, China
}

\affiliation{
	Hefei National Laboratory, Hefei 230088, China
}
\affiliation{
CAS Center for Excellence in Topological Quantum Computation and School of Physical Sciences,
University of Chinese Academy of Sciences, Beijing 100049, China
}
\author{Dongning Zheng}
\email{dzheng@iphy.ac.cn}
\affiliation{
	Beijing National Laboratory for Condensed Matter Physics,\\
	Institute of Physics,Chinese Academy of Sciences,Beijing 100190,China
}
\affiliation{
	School of Physical Sciences, University of Chinese Academy of Sciences, Beijing 100049, China
}
\affiliation{
	Hefei National Laboratory, Hefei 230088, China
}
\affiliation{
CAS Center for Excellence in Topological Quantum Computation and School of Physical Sciences,
University of Chinese Academy of Sciences, Beijing 100049, China
}	
\date{\today}

\begin{abstract}
A hybrid system with tunable coupling between phonons and qubits shows great potential for advancing quantum information processing. In this work, we demonstrate strong and tunable coupling between a surface acoustic wave (SAW) resonator and a transmon qubit based on galvanic-contact flip-chip technique. The coupling strength varies from $2\pi\times$7.0 MHz to -$2\pi\times$20.6 MHz, which is extracted from different vacuum Rabi oscillation frequencies. The phonon-induced ac Stark shift of the qubit at different coupling strengths is also shown. Our approach offers a good experimental platform for exploring quantum acoustics and hybrid systems.  
\end{abstract}

\maketitle
        \begin{figure*}[t]
		\includegraphics[scale=0.85]{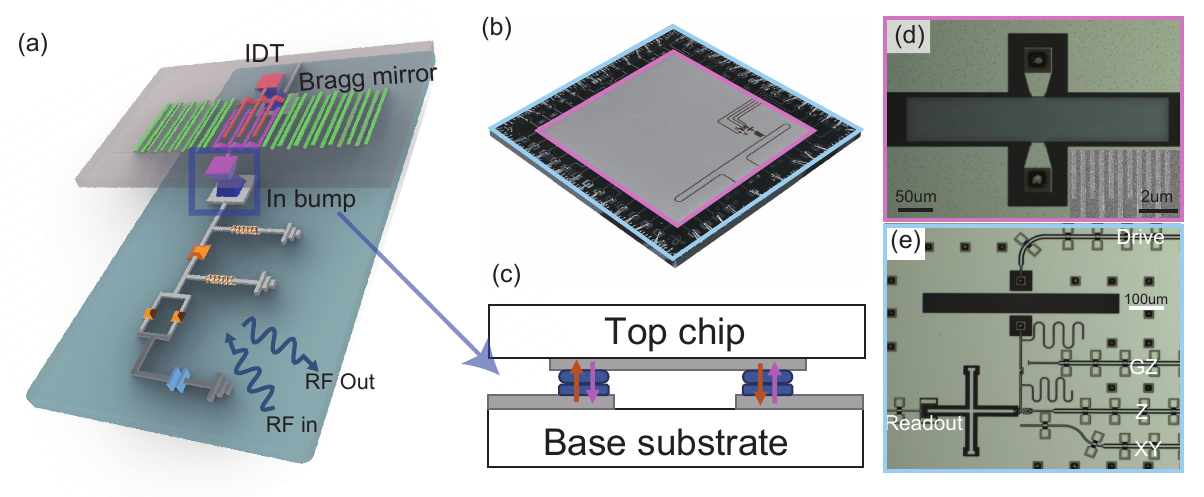}
		\caption{\quad The Device. (a) The 3D schematic of the sample. The SAW cavity consists of two Bragg mirrors and an IDT. One port of the IDT connects to the transmon quit with an indium bump and a gmon, while the other port connects to a driven line. The phonon modes of SAW are distributed on the surface of $\text{LiNbO}_3$. (b) Photograph of the device after flip-chip assembly. The base and top chip size is $15\,\text{mm}\times 15\,\text{mm}$ and $11\,\text{mm}\times 11\,\text{mm}$, respectively. (c) Side view of the assembled device. The indium (In) bumps of the top and base chip are compressed together under pressure and allow to realize the galvanic contact. (d) Optical micrograph of the top chip. (e) Optical micrograph of the base chip. The insert shows the scanning electron micrograph of zoomed stripe structures in the SAW cavity.}
        \label{figure1}
	\end{figure*}

Circuit quantum acoustodynamics (cQAD) studies the interaction between mechanical vibrations and superconducting qubits, which provides a way to control the non-classical mechanical modes (phonons).
There are various mechanical degrees of freedom including bulk wave~\cite{chu2017}, surface acoustic wave (SAW)~\cite{gustafsson2014,manenti2017a,bolgar2018,Moores_2018a,satzinger2018,Kitzman_2023a} and vibrations of suspended beams~\cite{oconnell2010,pirkkalainen2013,wollack2022,Lee_2023}. Among them, SAW is a traveling acoustic wave that propagates on the surface of the solid materials. Confining SAW between the Bragg mirrors can form the Fabry-Perot SAW cavity with achieved quality factor (Q) up to $100000$~\cite{manenti2016a,Emser_2022}. Compared to circuit quantum electrodynamics (cQED) systems~\cite{Blais_2021,Gu_2017}, acoustic waves travel five orders of magnitude slower than electromagnetic waves and the wavelength of gigahertz acoustic waves is close to that of optical waves. Therefore, the cQAD system has various applications, such as phonon-mediated quantum state transfer~\cite{bienfait2019}, quantum random access memory~\cite{hann2019} and medium for microwave-optical conversion~\cite{Forsch_2020,Mirhosseini_2020,Han_2021}. Some impressive work has been done to study the physical properties of cQAD systems, including preparation of phonon Fock states~\cite{satzinger2018,chu2018a}, splitting of phonons~\cite{Sletten_2019,Qiao_2023}, phonon induced non-exponential decay of giant atoms~\cite{Andersson_2019}, electromagnetically induced acoustic transparency~\cite{andersson2020}, squeezing and entanglement~\cite{wollack2022,Andersson_2022} of phonon states.
  
  For cQAD system, compared to weak coupling, strong coupling promises performing more coherent operations between superconducting qubits and other phononic resonators. Strong coupling regime can be achieved by improving the coherence time of the cQAD system or increasing the coupling strength. Fabricating the qubits directly on piezoelectric materials will induce phonon loss channel to the qubits. Therefore, it is better to fabricate the qubit and SAW on two different materials to keep good performance of cQAD system. There are two ways to realize the target. One way involves using piezoelectric films on sapphire substrate and placing the qubit on the area where the piezoeletric films are etched~\cite{jiang2023a,xu2022}. However, etching could cause damages to the substrate surface. The other option is to utilize flip-chip architecture to assemble two substrates made of different materials~\cite{chu2017,satzinger2018}. The reported flip-chip technology~\cite{satzinger2018,Kitzman_2023a} used in the cQAD system relies on glue. Therefore, the coupling strength between the qubit and SAW cavity is determined by the interchip geometric capacitance or mutual inductance and the grounding plane could be affected by glue bumps.

    Here, we report fabrication process and characterizations of the device which couples a GHz-frequency SAW resonator (cavity) to a transmon qubit. The qubit chip and SAW cavity chip are bounded together by flip-chip assembly with galvanic bumps (Indium). We verify the strong coupling between the SAW cavity and the qubit by performing vacuum Rabi oscillations experiment. Due to the existence of a tunable coupler, we can adjust the coupling strength in a range from $2\pi\times$7.0 to $-2\pi\times20.6\,$MHz. Then we measure the ac Stark shift of the qubit to show the effect of phonons inside the SAW cavity on the qubit. Our galvanic-connected flip-chip device can be realized with stronger coupling strength while ensuring the coherence of each elements.
    

    Our device, as shown in Fig.~{\ref{figure1}a}, is composed of two chips, which are bonded together using flip-chip technology. The top chip consists of a SAW resonator, fabricated on $128^{\circ}$Y-X lithium niobate ($\text{LiNbO}_3$) substrate -- a kind of material with strong piezoelectric effect. The SAW cavity is formed by an IDT (purple and orange) and two Bragg mirrors (green) on both sides. The IDT can conduct the conversion between acoustic waves and microwave, as well as excitation and detection of SAWs. The Bragg mirrors use grattings to form a Fabry-Perot cavity~\cite{manenti_2016}, which supports a stopband in the frequency response\cite{Morgan2007}. 
    The base substrate is sapphire and is used to fabricate qubits and control lines. The reason why we need two kinds of substrate is that the dielectric loss on $\text{LiNbO}_3$ is very large and will cause decoherence of qubits~\cite{Ioffe2007}.  The artificial atom we use is a frequency- tunable transmon qubit, formed by a DC SQUID and a shunted capacitor. The transmon qubit connects to the SAW cavity through a RF SQUID (i.e. gmon~\cite{chen2014,Geller2015}) with tunable inductance and indium bumps, which provides galvanic contact between the top and base chips. Compared to the designs in Ref.~\cite{satzinger2018}, we do not need to use interchip mutual inductance to make connection, which is sensitive to the space between the chips and has limited values. As show in Fig.~{\ref{figure1}e}, the transmon qubit is controlled by XY lines with the rotation in X or Y axis in the Bloch sphere and Z lines which tunes the transition frequency of qubit by applied flux bias. The state of the transmon qubit is read out by the dispersive-coupling coplanar waveguide resonator. 
    
    The fabrication of the device uses electron beam lithography (EBL) and optic laser lithography. First, we use ebeam evaporation to deposit $100\,$nm Al for the base chip and $30\,$nm Al for the top chip, respectively. Then, we use optic lithography and wet etch to form coplanar waveguides (CPWs) and ground planes. The stripes of the SAW cavity in the top chip are defined through EBL with width as $d=240\,$nm and length as $75\,$um using photoresist PMMA. These slender stripes are extremely difficult to be fabricated, especially when the periodic cell number of the Bragg mirrors at each side is $N_m=$400 and the periodic cell number of IDT is $N_t=20$. Before the evaporation process, we conduct the oxygen plasma cleaning in situ to remove the resist residue using \textit{Plassys}. The Al stripes are made by two steps. The first step is to create extra undercuts with the deposition angle at $\pm80^{\circ}$ and $30\,$nm Al, facilitating the lift-off process. The second step is to deposit $30\,$nm Al  at $0^{\circ}$ angle. This deposition method is also used in Ref.~\cite{bolgar2018}. The stripe structrues after lift-off are shown in Fig.~{\ref{figure1}d}. 
    The qubits in the base chip are exposed with EBL and fabricated with standard Dolan bridge shadow evaporation.
    The final step is to make flip-chip architecture using \textit{FC150} bonder. Although indium is a commonly used metal for flip-chip bonding~\cite{Datta2010} and becomes superconductive below $3.4\,$K, the direct contact between aluminum and indium will form an intermetallic~\cite{Foxen2018,Wade1975}, damaging the conductivity of these bumps. To address this issue, we use niobium as the diffusion barrier, which can be deposited via magnetron sputtering and also exhibits superconductivity below $9.2\,$K. Then, we deposit indium on each chips with the thickness of $7.5\,$\text{$\mu$}m using thermal evaporation. Note that we conduct ion mill in situ to make good electrical contact prior to each metal deposition. The two chips are bonded together with a force of $7000\,$g, ensuring the distance between them around $7.5\,$\text{$\mu$}m. The complete device is shown in Fig.~{\ref{figure1}b} with the side-view diagram in Fig.~{\ref{figure1}c}.

    
	\begin{figure}[]
		\includegraphics[scale=0.8]{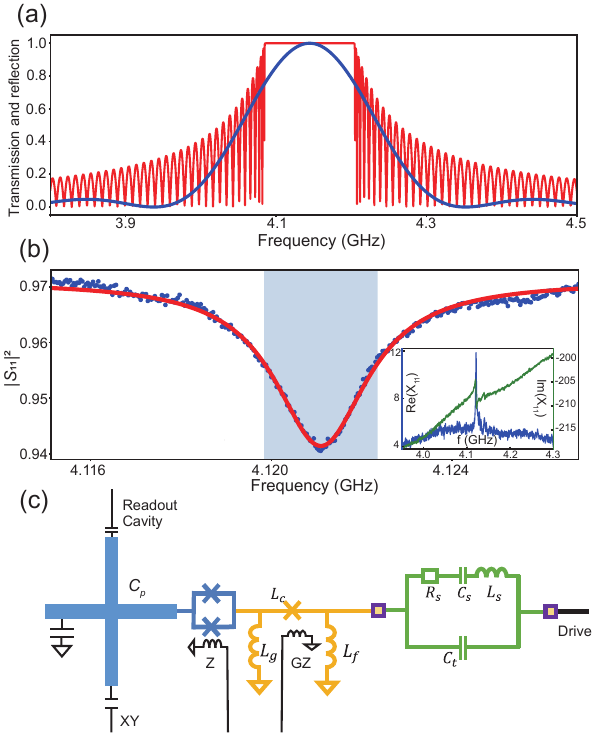}
		\caption{\quad Modelling of the SAW-Qubit coupling system. (a) The red line is calculated reflection of Bragg mirror \iffalse with reflection coefficient of single electrode $r_{s1}=0.045j$ .\fi and the blue line is the conductance of the IDT. \iffalse The corresponding bandwidths are $\Delta_{M}=120\,$MHz and $\Delta_{IDT}=184\,$MHz.\fi (b) The measured reflection of SAW cavity (blue dots) with Lorentizian fitting (red line). The insert shows the real and imaginary reactance $X_{11}$ of the SAW cavity, calculated from the measured \textit{S} data. The blue area shows the decay rates of the cavity as $\kappa_{S,RT}=2.5\,$MHz. (c) Circuit diagram of our SAW-Qubit coupling system. The green parts represent the equivalent circuit of SAW cavity, which is distributed in the top $\text{LiNbO}_3$ chip. The gmon coupler (Yellow) and the transmon qubit (blue) are in the base Sapphire chip.}
        \label{figure2}
	\end{figure}
 
    The frequency response of SAW cavity can be simulated with the coupling-of-mode (COM) model~\cite{Morgan2007}. The propagation speed of SAW in $128^{\circ}$Y-X $\text{LiNbO}_3$ is $v_a\approx3979\,$m/s at room temperature~\cite{Shibayama_1976}. The periodicity of IDT stripes is $p=4d$ and determines the central wavelength and frequency of the excited acoustic wave as $\lambda_0=p$ and $f_0=v_a/\lambda_0$. Both overexposure in EBL and the placed direction of IDT will affect its central frequency because $\text{LiNbO}_3$ is an anisotropic material. The calculated conductance of IDT is a $sinc$ function at $f_0$ with the bandwidth $\Delta_{IDT}=184\,$MHz, as shown in Fig.~{\ref{figure2}a}. The reflection of single electrode in the Bragg mirror is  $|r_{s1}|=0.045$. We need to conduct the Bragg mirror with a stripe array, in order to improve the total reflection rate to 1 by using the interference among them. The bandwidth of Bragg mirror is $\Delta_{M}=120\,$MHz, which should be less than $\Delta_{IDT}$, and is used to select frequency and improve Q factor. 
    
    We show the measured reflection data of the SAW cavity with $N_t=20$ and the distance between the two Bragg mirrors $L_0=20.5\lambda$ in Fig.~{\ref{figure2}b}. The equivalent inductance can be calculated with $L_s=0.5 \lim_{\omega\rightarrow\omega_0}\text{Im}[\partial X/\partial \omega]$, where the reactance $X$ in the insert can be calculated from the measured $S$ data and $\omega_0=2\pi f_0$.  The calculated BVD parameters of the SAW cavity in in Fig.~{\ref{figure2}b} are $L_s=186\,$nH, $C_s=\omega_0^2/L_s=8\,$fF and $R_s=Z_s/Q_s=\SI{2.92}{\Omega}$ with the characteristic impedance $Z_s=\sqrt{L_s/C_s}$ and the room temperature Q factor $Q_{S,R}=1643$. The electrostatic capacitance of IDT is $C_t=744\,$fF.

    \begin{figure*}[t]
    \centering
    \includegraphics[height=12cm]{"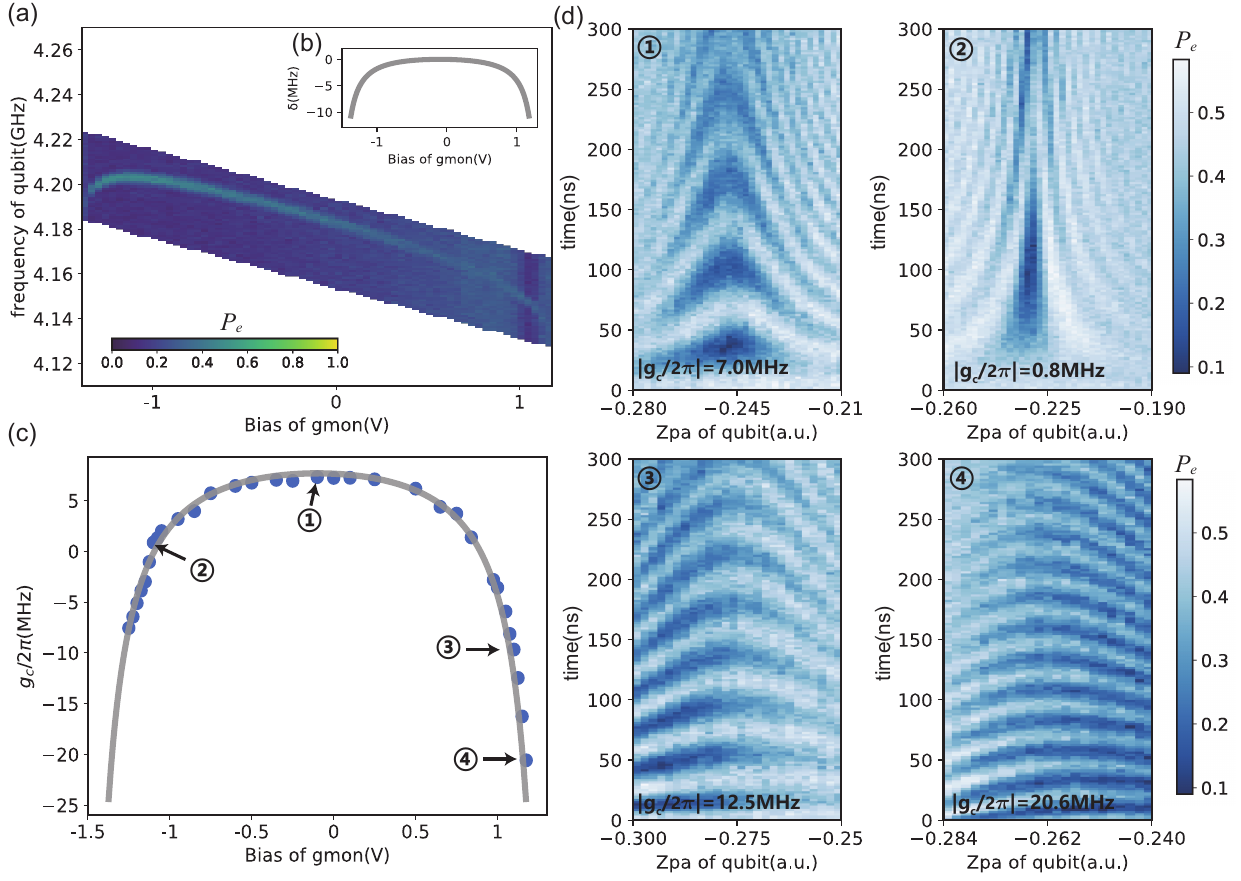"}
    \caption{\quad Characterization of tunable coupling between the SAW cavity and qubit via gmon. (a) The two-dimensional energy spectrum of qubit under different gmon biases. The fitting data after removing the influence of crosstalk is shown in (b). (c) The coupling strength of the SAW cavity and qubit under different gmon biases, and the curve morphology is consistent with (b). The blue dots are experimental data. (d) The time-domain oscillation data from points \ding{192}-\ding{195} in (c).}
    \label{figure3}
    \end{figure*}

  To study the cQAD system in the quantum regime, we mount the sample inside the mixing chamber of a dilution refrigerator with base temperature around $10\,$mk. The transmon qubit can be described by the charge energy $E_c/h=e^2/C_p=0.18\,$GHz with $C_p=108\,$fF and maximal Josephson energy $E_{j0}/h=22.5\,$GHz. The transition frequency from ground to first-excited state of the qubit can be turned by external flux $\Phi_e$ as $\omega_q=(\sqrt{E_c E_{j0}\cos (\Phi_e/\Phi_0)}-E_c)/\hbar$ with $\Phi_0=h/2e$. The Hamiltonian of the system shown in Fig.~{\ref{figure2}c} can be described by Jaynes-Cummings model as
    \begin{align}\label{Hs}
    H=\omega_{0}' a^{\dag} a+\frac{\omega_q}{2} \sigma_z+g_c\left(a^{\dag} \sigma_{-}+a \sigma_{+}\right),
    \end{align}
    where $a$ is the annihilation operator of SAW cavity mode, $\sigma_{-}$ ($\sigma_{+}$) is the annihilation (creation) operator and $\sigma_z$ is the population operator of the transmon qubit. $\omega_{0}'$ is the resonance frequency of SAW cavity at low temperature. $g_c$ is the coupling strength between SAW cavity and qubit and is adjustable as
      \begin{align}\label{gc}
      g_c=\frac{1}{2}\frac{1}{\sqrt{\left( {{L}_{J}}+{{L}_{g}} \right)\left( {{L}_{s}}+{{L}_{f}} \right)}}\frac{{{L}_{g}}{{L}_{f}}}{{{L}_{g}}+{{L}_{f}}+{{L}_{c}}}{{\omega }_{0}'},
      \end{align}  
      where ${L}_{c}={{L}_{c0}}/\cos \delta$ is the tunable inductance of gmon and $\delta$ is the phase difference across the junction, tuned by the DC bias in the RF SQUID. The parameters of the gmon loop are $L_g=475\,$pH, $L_f=523\,$pH and $L_{c0}=645\,$pH, which are obtained by the measured coupling strengths (discussed later). $L_J$ is the inductance of the transmon qubit, which is around $10\,$nH and can be tuned by the external flux. 

Figure{~\ref{figure3}a} shows how the transition frequency of qubit is modulated by the bias of gmon. The applied voltage converts to biased current in the local control lines on chip because of the series resistors in the input lines.  The effective inductance of gmon changes with the bias voltage and affects the transition frequency of qubit. Moreover, the applied voltage to gmon causes Z-crosstalk to the qubit. After correcting, the modulated curve is shown in Fig.~{\ref{figure3}b}.

We prepare the qubit to excited state $|e\rangle$ (denote ground and excited state of qubit as $|g\rangle$ and $|e\rangle$, the Fock state of SAW cavity as $|0\rangle$, $|1\rangle$) then track the time-domain response of the qubit over a narrow range around the frequency of SAW cavity by qubit fast Z-control pulse. As the qubit bias is moved to resonance with SAW cavity causing oscillation between dressed states $|0e\rangle$ and $|1g\rangle$, we expect the beating frequency to decrease to $2g_{c}$ as shown in Fig.{~\ref{figure3}d}. The frequency of SAW cavity can be obtained as ${\omega }_{0}'/2\pi = 3.901\,$GHz. We conduct experiments under different gmon biases in Fig.~{\ref{figure3}c} to investigate the variation of qubit-SAW cavity coupling strength with gmon bias. Fig.~{\ref{figure3}d{\ding{192}}}-{\ding{195}} presents the Rabi oscillations of different effective coupling strength obtained in our measurements, Fig.~{\ref{figure3}d\ding{195}} shows a maximum $g_c/2\pi = 20.6\,$MHz.

\begin{figure*}[t]
\centering
\includegraphics[height=4.5cm]{"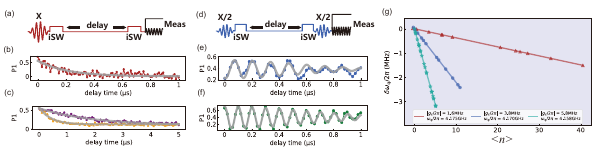"}
\caption{\quad The characterization of the phonon in SAW cavity. (a) The pulse sequence for the measurement of the phonons' energy relaxation time $T_{1S}$. X means the X gate for qubit. (b) The red dots represent the data obtained from experimental measurements, while the gray curve represents the fitted data. (c) Decoherence time of qubit at different frequencies. The $T_1$ of the qubit at the swap point is illustrated with orange points, measured as $T_{1q}^{swap} \approx 1881\,$ns  when the coupling is turned off. The $T_1$ of the qubit at the idle point is shown with purple points, measured as $T_{1q}^{idle} \approx 452\,$ns. (d) The pulse sequence for the measurement of the phonons' dephasing time $T_{2S}$. X/2 means the X/2 gate for qubit. (e) The measured data of $T_{2S}$ is illustrated with blue points and the gray curve is fitting data. (f) The measured data of $T_{2q}$ at idle frequency is illustrated with green points and the gray curve is fitting data. (g) Characterize the number of phonons in the SAW cavity using the ac Stark effect at different coupling strengths. }
\label{figure4}
\end{figure*}

The pulse sequence for the measurement of the phonon's energy relaxation time ($T_{1S}$) in SAW cavity is depicted in Fig.~{\ref{figure4}a}. The qubit is excited with X gate while off-resonance, and then coupled to the SAW cavity for $40\,$ns to transfer the state with an iSWAP. A second iSWAP 
later retrieves the state back to qubit for readout. The resulting decay in Fig.~{\ref{figure4}b} gives the lifetime of a single phonon in SAW cavity yields $T_{1S}^{Meas} \approx 205\,$ns.

The entire relaxation process can be divided into three parts: the first iSWAP gate, the delay and the second iSWAP gate, as denoted in Fig.~{\ref{figure4}a}. The dissipation experienced by the system during the swap process is contributed by both the qubit and the cavity, depending on their individual dissipation rates $\gamma_q^{swap}$ and $\gamma_S$, respectively. On the other hand, the qubit and SAW cavity are in a large detuning during the delay, introducing a Purcell effect $\gamma_S^{*} = \gamma_S + ({g_c}/{\Delta})^2\gamma_q^{idle}$ for SAW cavity, where $\Delta = \omega_q - {\omega }_{0}'$ and $\gamma_q^{idle}$ is the dissipation rates of qubit at idle frequency. Two iSWAP gates only change the initial value of the curve so the population of SAW cavity being in the entire process is denoted as 
\begin{eqnarray}
	P_1^{SAW}(t) = P_0\cdot e^{\left\{-t\cdot [\gamma_S +(g_c/\Delta)^2\gamma_q^{\text{idle} }]\right\}}.
\end{eqnarray}

According to the experimental parameters in Fig.~{\ref{figure4}b} and Fig.~{\ref{figure4}d}, the additional dissipation caused by the Purcell effect is much smaller than intrinsic dissipation, i.e. $(g_c/\Delta)^2\gamma_q^{idle }\ll\gamma_S$. Therefore, the $T_{1S}$ we measured is reliable which allows the determination of the quality factor $Q^{SAW} = \omega_{0}' \cdot T_{1S} \approx 5025$.The $T_2$ of SAW cavity can be measured by the pulse sequence shown in Fig.~{\ref{figure4}d} which is similar to the case of $T_1$. Figs.~{\ref{figure4}e} and {\ref{figure4}f} show the measured data of SAW cavity and the qubit at the idle point, respectively.

Finally, we used the ac Stark effect to calibrate the phonon number in the SAW cavity. According to the results of c-QED in dispersive regime, the frequency of the qubit varies with the number of phonons in the cavity
\begin{eqnarray}
	\delta\omega_q \equiv \omega_q' - \omega_q = 2\chi n, \chi = \frac{g_c^2\eta}{\Delta(\Delta+\eta)}
	\label{chi}.
\end{eqnarray}

Applying resonance driving with different powers to the SAW cavity and scanning the spectrum of qubit at the same time. According to $\eta/2\pi = -180\,$MHz, we calibrated the average number of phonons $\langle n \rangle $ by frequency shift of qubit as $\langle n \rangle = \delta\omega_q/2\chi$. We adjusted the gmon's bias and measured the qubit frequency shift $\delta\omega_q$ at different coupling strengths, as shown in Fig.~{\ref{figure4}g}. According to Eq.~\ref{chi}, with the same $\Delta$ and $\eta$, the slope of $\delta\omega_q$ versus the $\langle n \rangle $ should be proportional to $g_c^2$. This is consistent with our experimental results. It also reveals another method for measuring electro-acoustic coupling.

In summary, we integrate the SAW cavity and the transmon qubit together using flip-chip assembly with indium bumps. Using this cQAD device, we observe the strong coupling between the qubit and SAW cavity and show the tunability of the coupling strength. Furthermore, we observe the ac stark shift of the qubit causing by the phonons in the SAW cavity. Comparing the characterization of SAW cavity in room temperature, we observe an improvement in the Q factor under lower temperatures. These can be explained by the smaller acoustic dissipation  of $10\,$mK within the high-vacuum environment. Our work provides a new platform for studying cQAD and a new way to integrate them. 
The galvanic-connected flip-chip technology gives a modular way to optimize the qubit and mechnical modes separately. This new platform has potential applications in quantum communication and quantum computing.

\begin{acknowledgments}
This work was supported by the National Natural Science Foundation of China (Grants No. 12204528, No. 92265207, No. T2121001, No. 11934018, No. 12005155, No. 11904393, No. 92065114, No. 12204528, No. 11875220, and No. 12047502), Innovation Program for Quantum Science and Technology (Grant No. 2021ZD0301800), and Scientific Instrument Developing Project of Chinese Academy of Sciences (Grant No. YJKYYQ20200041). Z.H. Peng also acknowledges the funding support by the National Natural Science Foundation of China (Grant No. 12074117, No. 92365209). J. Zhang also acknowledges the funding support by Tsinghua-Foshan Innovation Special Fund (TFISF), Innovative leading talent project of "Shuangqian plan" in Jiangxi Province, Joint Fund of Science and Technology Department of Liaoning Province and State Key Laboratory of Robotics, China (2021-KF-22-01) and Guoqiang Research Institute of Tsinghua University (20212000704). X.H. Ruan thanks for the measurement support from Xu Wang in CAS. This work also was supported by the Micro/nano Fabrication Laboratory of Synergetic Extreme Condition User Facility (SECUF). Devices were made at the Nanofabrication Facilities at the Institute of Physics, CAS in Beijing.
\end{acknowledgments}

\section*{AUTHOR DECLARATIONS}
\subsection*{Conflict of Interest}
The authors have no conflicts to disclose.

\bibliography{refrence}

\end{document}